\documentclass[12pt]{article}
\usepackage{epsf}
\usepackage{graphicx}

\newlength{\dinwidth}
\newlength{\dinmargin}
\def\lapproxeq{\lower .7ex\hbox{$\;\stackrel{\textstyle <}{\sim}\;$}}
\def\gapproxeq{\lower .7ex\hbox{$\;\stackrel{\textstyle >}{\sim}\;$}}

\def\be{\begin{equation}}
\def\ee{\end{equation}}
\def\bea{\begin{eqnarray}}
\def\eea{\end{eqnarray}}

\begin{document}
\titlepage

\begin{flushright}
IPPP/09/50\\
DCPT/09/100\\
\today \\ 
\end{flushright}

\vspace*{4cm}

\begin{center}
{\Large \bf Radiative QCD backgrounds to exclusive \\[2mm]
$H\to b \bar b$ production: radiation
from the screening gluon}

\vspace*{1cm} \textsc{V.A.~Khoze$^{a,b}$, M.G. Ryskin$^{a,b}$ and A.D.Martin$^{a}$} \\

\vspace*{0.5cm} $^a$ Department of Physics and Institute for
Particle Physics Phenomenology, \\
University of Durham, DH1 3LE, UK \\[0.5ex]
$^b$ Petersburg Nuclear Physics Institute, Gatchina,
St.~Petersburg, 188300, Russia \\[0.5ex]
\end{center}

\vspace*{1cm}

\begin{abstract}

Central exclusive Higgs boson production, $pp\to p \oplus H \oplus p$, at the LHC
can provide an important complementary contribution
to the comprehensive
study of the Higgs sector
in a remarkably clean topology.
The  $b \bar b$ Higgs decay mode is especially attractive, and for certain BSM scenarios may 
even become {\it the} discovery channel. 
Obvious requirements for the success
of such exclusive measurements are strongly suppressed and controllable backgrounds. 
One potential source of background comes from additional gluon radiation
which leads to a three-jet  $b \bar b g$ final state.
We perform an explicit calculation of the 
subprocesses  $gg\to q\bar q g$,  $gg\to ggg$
in the case of  `internal' gluon radiation from the spectator, $t$-channel
screening gluon, when the two
incoming active $t$-channel gluons form a colour octet.
We find that the overall contribution of this source of background 
is orders of magnitude
 lower than that caused by the main irreducible
background resulting from the $gg^{PP}\to b\bar b$ subprocess.
Therefore, this background contribution can be safely neglected.

\end{abstract}

\newpage

\section{Introduction \label{sec:intro}}

The search for the Higgs boson(s) is one of the main goals of the ATLAS and CMS
experiments at the LHC. 
Once the Higgs boson is discovered, it will be of primary importance
to determine its spin and parity, and to measure precisely
the mass, width and couplings. A comprehensive study of the whole Higgs sector,
including precision  mass and coupling measurements, spin and CP properties, will be the next stage.
Measurements of various important
properties of the Higgs sector may be very challenging for the
traditional LHC searches. In particular, the direct
determination of the $b \bar b $ Yukawa coupling
is inaccessible  due to overwhelming QCD backgrounds. Moreover,
nearly degenerate Higgs-like states are extremely 
difficult to separate. The spin-parity identification
in many popular `Beyond the Standard Model' (BSM) scenarios, as well as an observation 
of possible `invisible' Higgs decays, exemplify other challenging
issues.  

The conventional strategy to achieve the ambitious programme of
a complete study of the properties of the Higgs sector
requires an intensive interplay between the LHC and the ILC (high-energy linear $e^+e^-$ collider),
see for example, \cite{ILC}.
Whilst awaiting the possible arrival of the ILC, there has been a growing interest in recent years
in the possibility to complement the standard LHC physics menu by 
installing  near-beam proton detectors in the LHC tunnel some 420 m from the
ATLAS and CMS interaction points \cite{FP420,pb}.
In this way we may address many of the  challenging issues well before the ILC has become
operational, see, for example, \cite{KMRprosp}~-~\cite{katri}
and references therein. 

Although experimentally difficult, 
the forward proton mode at the LHC 
would provide an exceptionally clean environment to search for, and to
identify the nature of, new objects, in particular, Higgs-like bosons.
The expected cross sections are very small (about four orders of magnitude
lower than those for the standard inclusive processes).
Thus, an obvious requirement for the success
of such measurements is that the  backgrounds should be strongly suppressed
and controllable.

Recall that, though the 
expected total
production cross section for the SM Higgs at the LHC is rather large (around 50 pb),
in order to identify the signal in the quite hostile
background environment we have to rely, either on
 rare decay
 modes such as $H \to\gamma \gamma$, or to impose very
severe cuts on the final state configurations\footnote{
In this respect
a selection of the quasi-three body $pHp$ final state 
could be viewed  as an extreme example of a final state cut.}.
This unavoidably leads to a strong reduction (typically, by 3-4 orders
of magnitude) of the observed cross sections.
In this work we revisit the evaluation of the QCD backgrounds to exclusive 
 $H\to b \bar b$ production at the LHC. 
Previous studies can be found in \cite{DKMOR,hkrstw}, \cite{krs1}~-~\cite{Blois07}.
However, the main emphasis here is on radiation off the so-called screening gluon in the basic QCD diagram for
exclusive $ b\bar b$ production. This has not been quantified before.

 The structure of the paper is as follows. 
In Section 2  we recall the formalism used to calculate the cross section of central exclusive Higgs boson production. Then in Section 3  the main sources of background to the $pp\to p+(H\to b\bar b)+p$ signal are considered. In Section 4 we describe the structure of the amplitude of internal gluon radiation from the screening gluon, while in Sections 5 and 6, respectively, the cross sections for the hard subprocesses of $b\bar b$ and $gg$ dijet production in a colour-octet state are calculated. The numerical results are presented in Section 7, and a brief summary given in Section 8.

\section{Central Exclusive Diffractive Higgs Production  \label{sec:cep}}

The central exclusive production (CEP) of a Higgs boson is the process
$pp\to p \oplus H \oplus p$, where the $\oplus$ denote the 
presence of large rapidity gaps
 between the outgoing protons and the decay products
of the central system. As already mentioned, 
CEP offers
 a unique complementary measurement to the
 conventional Higgs search channels. 
First, if the outgoing protons scatter
through small
angles, then, to a very good approximation, the primary active di-gluon
system obeys a $J_z=0$, $CP$-even selection rule~\cite{KMRmm}. 
Here $J_z$ is the projection of the total
angular momentum along the proton beam axis.
The observation of the Higgs boson in the CEP
channel, therefore, determines the Higgs quantum numbers
to be dominantly $J^{PC}=0^{++}$, see \cite{KKMRext,KKMRcentr} for details. 
Secondly, because the process is exclusive, 
 all of the energy/momentum lost by the protons during the
 interaction goes into the production of the central system.
 Measuring the outgoing protons allows the central mass, $M_H$, 
to be 
 determined with an accuracy of just a few GeV 
 regardless of the decay products of the central system. At the
same time, the equality of the accurate missing-mass reconstruction 
of the Higgs, 
$M_H$, with its mass determined in the central detector
 from the decay products allows the background to be considerably suppressed.
A precision missing-mass measurement  requires dedicated 
forward proton detectors to be installed 
in the high dispersion region
 420 m either side of the ATLAS and/or CMS interaction points\footnote{We refer the reader to the FP420 R\&D report for experimental
and detector 
details \cite{FP420}.}. 
It is important to note that a signal-to-background ratio of order 1
(or even better) is achievable \cite{KMRmm,DKMOR}.
As discussed in \cite {hkrstw,hkrtw}, CEP processes
would enable a unique signature for the MSSM 
Higgs sector, in particular allowing the direct measurement of the $Hbb$ Yukawa 
coupling.
 Furthermore, CEP can provide valuable information on the
 Higgs sector of NMSSM \cite{fghpp} and other popular
BSM scenarios, and can, for example, be also  beneficial in searches for 
Higgs triplets \cite {katri}.

Finally, in some BSM schemes this mechanism
provides an opportunity for lineshape analysis \cite{JE,KKMRext},
including the Higgs width measurement in the case
of large width, and allows the direct observation of a $CP$-violating signal \cite{JE,KMRCP}.
As already mentioned, the final-state structure is much cleaner
than in the (messy) non-diffractive environment, and the event kinematics are strongly
constrained by measuring the outgoing protons. Moreover, the study of the
azimuthal correlations between the final protons would allow a straightforward  
approach to 
probe the $CP$ structure of the  Higgs sector \cite{KMRCP}.
\begin{figure}
\begin{center}
\includegraphics[height=5cm]{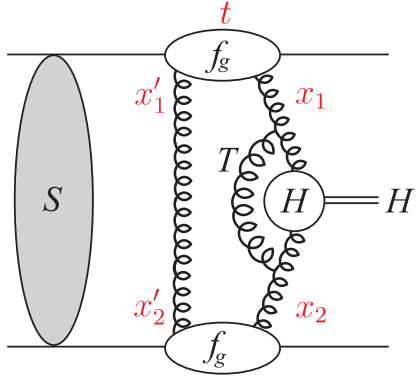}
\caption{A symbolic diagram for the central exclusive production of a Higgs boson $H$.}
\label{fig:parts}
\end{center}
\end{figure}

The theoretical formalism \cite{KMR}~-~\cite{MRKepip}
for central exclusive production contains distinct parts, as illustrated
in Fig.~\ref{fig:parts}. The cross section can be written schematically
in the form \cite{KMRprosp,KMR}
\begin{equation}
\sigma (pp \to p+H+p) ~\sim~\frac{\langle S^2 \rangle}{B^2} \left| \, N\int \frac{dQ_t^2}{Q^4_t}f_g(x_1,x'_1,Q^2_t,\mu^2)f_g(x_2,x'_2,Q^2_t,\mu^2) \right|^2
\label{eq:d1}
\end{equation}
where $B/2$ is the $t$-slope of the proton-Pomeron vertex,
 $\langle S^2 \rangle$ is the so-called
 survival probability of the rapidity gaps,
and the normalization, $N$, is given in terms of the $H\rightarrow gg$ decay width.
The amplitude-squared factor, $\left|...\right|^2$, can be calculated 
in perturbative QCD because the dominant contribution 
 to the integral comes from the 
 region $\Lambda^2_{\rm QCD} \ll Q^2_t \ll M^2_H$ 
 for the 
 large Higgs mass values of interest \cite{KMR}. 
 The probability amplitudes, 
 $f_g$, to find the appropriate pairs of $t$-channel gluons 
 $(x_1,x'_1)$ and $(x_2,x'_2)$ are given
  by skewed unintegrated gluon densities at a hard scale $\mu \sim M_H/2$.
 These generalized gluon distributions
 are usually taken at $p_t=0$, and then the ``total'' exclusive 
 cross section is calculated by integrating over the transverse 
 momentum, $p_T$, of the recoil protons. If we assume an 
 exponential behaviour then, using  
\be
\int dp^2_T~e^{-Bp_T^2}~=~1/B~=~\langle p_T^2\rangle,
\ee
we see that the additional factor in (\ref{eq:d1})
is  $\langle S^2 \rangle/B^2$
\cite{KMRprosp,MRKepip,KMRS}, which
has the form  $S^2\langle p^2_T \rangle^2$, and is much less dependent on
 the parameters of the soft model \cite{KMRnns,MRKepip,KMRS} than $S^2$ on its own.
The CEP cross section for Higgs bosons produced by gluon-gluon fusion, and decaying to $b\bar{b}$, is proportional to
\begin{equation}
\frac{\Gamma^{\rm eff}}{M_H^3} \equiv \frac{\Gamma(H\rightarrow gg)}{M_H^3} {\rm BR}(H\rightarrow b\bar{b})
\end{equation}
where $\Gamma(H\rightarrow gg)$ is the decay width to gluons and BR($H\rightarrow b\bar{b}$) is the branching ratio to $b\bar{b}$ quarks.
Comprehensive recent studies performed in \cite{KMRnns} show that
the effective value of the survival factor (which
accounts for the so-called `enhanced absorptive corrections' 
and other effects violating soft-hard factorization) is
\be
\langle S^2_{\rm eff} \rangle~=~0.015^{~+0.01}_{~-0.005}
\label{eq:s}
\ee
for CEP of a SM Higgs of mass around $M=120$~GeV.
Moreover, as discussed in \cite{MRKepip}, this result 
should be regarded rather as a conservative (lower) limit for the gap survival.
For the SM Higgs width  $\Gamma(H \to gg)=0.25$~MeV the resulting value
for the exclusive cross section is, 
 conservatively,
\be
\sigma(pp \to p+(H\rightarrow b\bar{b})+p)~\simeq~2~{\rm fb},
\label{eq:x}
\ee
with an overall uncertainty
of a factor of 3 up or down, see also \cite{KKMRext,KMRearly}
\footnote{Recall that for
 central exclusive  $ b\bar b$ production this uncertainty does not affect
the physically important signal-to-background ratio, $S/B$.}. 
The good news is that exclusive dijet \cite{CDFdijet},
 $\gamma \gamma$ \cite{CDFgg} and $\chi_c$ \cite{CDFchi} production data from CDF
 and the leading neutron data at HERA are in
a broad agreement \cite{MRKepip,JHEP,vakjj} with the basic theoretical
formalism of Refs.\cite{KMRprosp,KMR,KMRsoft} indicating
that it is unlikely that enhanced absorption effects will strongly
reduce the exclusive SM Higgs signal at the LHC energy. 

\section{On the backgrounds to the $p+(H\to b\bar b)+p$ signal \label{sec:bgd}}

The potential importance of the $p+(H\to b\bar b)+p$ process
means that the physical backgrounds to this reaction
must be thoroughly addressed.
These backgrounds can be broken down into three
broad categories: central exclusive, interaction
of two soft Pomerons and
the so-called overlap background.
The last source of background is important when there are a large number
of $pp$ interactions in each bunch crossing at the LHC. The largest
overlap background is a three-fold coincidence between two soft
single diffractive
 events ($pp\rightarrow pX$), which produce forward protons within
the acceptance of the forward detectors, together with an inelastic event, which
produces the hard scatter $pp\rightarrow b\bar{b}$ and,
thus, can mimic the signal. This 
important background is the subject of 
detailed studies \cite{FP420}, 
and it is shown that, with dedicated fast-timing detectors, and
some additional experimental cuts, it can be reduced to a tolerable level,
see, for instance \cite{cox1,katri}.
We do not consider this type of background  further.

Double-Pomeron-exchange (DPE) is the process
$pp \rightarrow p + X + p$, where the central system, $X$, is
produced by an inelastic Pomeron-Pomeron interaction.
In this case 
there are
always `Pomeron remnants' accompanying the hard scatter.
The DPE $b\bar{b}$ background has been extensively studied in relation to the
$H\rightarrow b\bar{b}$ signal, and it has been concluded that this 
background is negligible after appropriate experimental cuts
\cite{cox1,KMRins}.
We do
not consider these types of background events further as well.

Here, our concern is the exclusive $b\bar{b}$ background processes, which
are generated by the
collisions of the two hard (active) gluons in Fig.~\ref{fig:parts}, labelled $x_1, x_2$. 
The dominant sources of such background
were discussed in detail in \cite{DKMOR,hkrstw,krs1,shuv}.
It was shown that all these backgrounds are strongly  suppressed and controllable and,
in principle, can be
further reduced by the appropriate optimized cuts on the final state particle
configurations in such a way that the signal-to-background ratio, $S/B$, is
of order 1
(or may be even better for MSSM 
or other BSM schemes \cite{hkrstw}~-~\cite{KKMRext}, \cite{katri}).

The unique 
advantage of the $p+(H\to b\bar b)+p$
signal is that there exists
a $J_z=0$ selection rule \cite{KMRmm,Liverpool}, which requires
the leading order (LO)
$gg^{PP}\to b\bar b$ background subprocess to vanish in the limit of
massless quarks and forward outgoing protons. Here
the $PP$ superscript is to indicate that
that each active gluon
comes from colour-singlet $t$-channel (Pomeron) exchange.
However, there are still four main sources of
$gg$-generated backgrounds, some of which occur even at LO \cite{hkrstw,krs1,Blois07}.
\begin{itemize}
\item[(i)] 
The prolific (LO)
$gg^{PP}\to gg$ subprocess can mimic $b\bar b$ production since we may
misidentify the  gluons as $b$ and $\bar{b}$ jets\footnote{
When calculating the cross section of exclusive 
gluon-dijet production we neglect the diagrams shown in Fig. 7 and Fig. 8c of~\cite{cudell}.
 The contribution of the diagram of Fig. 7
 is power-suppressed by a factor 
of $Q^2_t/E^2_T$, where $E_T$ is the transverse energy of the jets. For LHC energies,
and $M_H=120$ GeV, the main contribution to the exclusive amplitude comes from the region of $Q_t\sim 2$ GeV.
The small 
contribution from the $Q_t\sim E_T$ domain can be considered as a minor part of the NLO correction to the LO result.
 The diagram of  Fig. 8c in \cite{cudell} is treated as part of the Sudakov form factor. However, this diagram 
does not allow for double logarithms. Even the single-log contribution is questionable here. Recall that, actually, we calculate the {\it imaginary} part, that is the $s$-channel discontinuity, of the CEP amplitude. If required
 the numerically small (for our positive signature case) real part can be restored by including the well known signature factors. The only possible logarithm in the $s$-channel discontinuity of Fig. 8c is the BFKL-type longitudinal logarithm, which can be removed by choosing the appropriate 
planar gauge, where the gluon field $A_\mu$ is orthogonal to the vector $n_\mu$ ($A_\mu n_\mu=0$), say with  $n=(p_1+p_2)_\|$ parallel to the longitudinal part of the dijet momentum $(p_1+p_2)$. In a different gauge the  diagram of  Fig. 8c
may lead to a longitudinal single-log, coming from the region of  relatively low gluon transverse momentum $q_t \ll E_T$. This piece should be summed up, together with other analogous terms caused by permutations of the gluon lines. As  discussed in~\cite{KMRearly} the contribution from the region of very
 small $q_t<Q_t$ vanishes due to destructive 
interference between the emission from the active gluon ($x$) and the screening gluon. To trace this cancellation explicitly we have to consider  {\it all}  permutations, which at the single-log accuracy are summed up by the LO BFKL kernel. Using the known BFKL kernel it was shown in \cite{KMRearly} (Sect. 5) that the sum of all  permutations leads to the effective
 lower cutoff $q_t=Q_t$ in the integral for the Sudakov-like form factor $T$. This value of the cutoff provides the single-log accuracy of the $T$-factor. The NLO contribution
of Fig.8c, being not enhanced by any large logarithm, is the part of the NLO corrections, which are beyond the accuracy of our approach.}.
\item[(ii)]
An admixture of $|J_z|=2$ production, arising when 
the outgoing
protons have non-zero $p_t$, which contributes to the
LO $gg^{PP}\to b\bar b$ background. 
 
\item[(iii)] 
Because of the non-zero mass of the $b$ quark there is a contribution to the $J_z=0$ 
$gg^{PP}\to b\bar b$ cross section of order $m_b^2/E_T^2$.
Here  $E_T$ is the
transverse energy of the $b$ and $\bar{b}$ jets.

\item[(iv)] 
There is a possibility of NLO $gg^{PP}\to b\bar b g$
background contributions, which for large angle, hard
gluon radiation do not obey the selection rule. 
In particular,
the extra gluon may go unobserved in the direction of
a forward proton. This background is reduced by
requiring the approximate equality $M_{\rm missing} = M_{b\bar b}$.
Calculations \cite{KMRins} show that this  background  
 may be safely neglected.
The remaining danger is large-angle hard gluon emission, which is
collinear with either the $b$ or $\bar{b}$ jet, and, therefore,
unobservable. 
This background source results
in a sizeable contribution, see \cite{krs1}, which,
together with other pieces listed in items (i-iii),
are accounted for in the evaluation \cite{hkrstw,hkrtw} of the prospective sensitivities
for the exclusive production of scalar Higgs bosons at the LHC. 
\end{itemize}

Among all the QCD backgrounds, the $m_b^2/E_T^2$ -suppressed dijet $ b \bar b$ production
is especially critical, since it is practically the only irreducible
background source, which cannot be decreased, either by improving the hardware
(as in the case, when the prolific
$gg^{PP}\to gg$ subprocess  mimics $ b \bar b$
 production, with the outgoing gluons misidentified as
 $b$ and $\bar{b}$ jets \cite {DKMOR})
or, for example, by cuts on the three-jet event topology
(as in the case of large-angle gluon radiation in the process
$gg (J_z=0)\to q\bar q g$, discussed in \cite{hkrstw,cox1,krs1}).
For some time the $m_b^2/E_T^2$ -suppressed
term 
raised concern, since the result could be
 strongly affected by large higher-order QCD effects. The good news is,
that, as shown in Ref. \cite{shuv},
the one-loop
QCD  corrections suppress the exclusive
$b\bar b$ background (by a factor of about  2,
or more for larger $b\bar b$-masses), 
in comparison with that calculated using the Born $gg^{PP}\to b\bar b$
amplitude. This result has been already accounted for in
the numerical studies \cite{hkrtw} of CEP  of MSSM Higgs bosons.

 The   $b\bar b$
production in the $|J_z|=2$ state is
another sizeable  background. In  principle,
this contribution can be
reduced by selecting events with forward outgoing protons of smaller transverse momenta \cite{KMRmm}. However,
at the moment, among the background sources listed
in items(i-iv) above  the exclusive gluon-gluon dijet production is one
of the most
important backgrounds for the CEP Higgs process.
In order to
suppress further this QCD contribution we need better experimental
 discrimination between
$b$-quark and gluon jets; that is, to achieve a lower probability
$P_{g/b}$ for misidentifying a gluon as a $b$-jet.

Let us elucidate the physical origin of the
suppression of the LO
$gg^{PP}\to b\bar b$  subprocess. 
As discussed in Ref. \cite{krs1},
it is convenient to consider
separately the  quark helicity conserving
(QHC) and the quark helicity non-conserving (QHNC) 
background amplitudes. As shown in Refs. \cite{krs1,borden,fkm},
this suppression is a direct consequence
of the symmetry properties of the Born helicity amplitudes,
$M_{\lambda_1,
\lambda_2}^{\lambda_q,\lambda_{\bar q}}$, describing the binary background process
\begin{equation}
g(\lambda_1, p_A) \: + \: g (\lambda_2, p_B) \;
\rightarrow \; q
(\lambda_q, p_1) \: + \: \overline{q} (\lambda_{\bar q}, p_2) \/ .
\label{eq:a1}
\end{equation}
Here, the $\lambda_i$ label the helicities of the incoming gluons, and
$\lambda_q$ and $\lambda_{\bar q}$ are the (doubled) helicities of the
produced quark
and antiquark.  The  $p$'s denote the particle
four-momenta ($p_A^2=p_B^2=0$, $p_{1,2}^2=m^2$), with
$p_A+p_B=p_1+p_2$ and $s=(p_A+p_B)^2$.
For a colour-singlet,  $J_z = 0$, initial state,
$(\lambda_1=\lambda_2\equiv \lambda)$
the Born QHC amplitude
with $\lambda_{\bar q} = -\lambda_q$ vanishes
\cite{fkm} \footnote{Note that in the massless limit
Eq.~(\ref{eq:a2}) holds for any colour state of initial gluons.
This is a consequence of the general property, that the non-zero massless tree-level
amplitudes should contain at least two positive or two negative helicity
states, see, for example, \cite{mhv1}. It is a
particular case of the more general Maximally-Helicity-Violating
amplitude (MHV) rule, see, for example, \cite{MP}.}
\begin{equation}
 M_{\lambda, \lambda}^{\lambda_q, -\lambda_q} \; = \; 0.
\label{eq:a2}
\end{equation}
For the QHNC amplitude for
large-angle
production we have
\begin{equation}
M_{\lambda, \lambda}^{\lambda_q, \lambda_q} \; \sim \; {\cal O} \left (
\frac{m_q}{\sqrt{s}}
\right ) \: M_{\lambda, - \lambda}^{\lambda_q, -\lambda_q},
\label{eq:a3}
\end{equation}
where the amplitude on the right-hand-side displays the dominant
helicity configuration of the LO background process.
The above-mentioned $m_b^2$-suppression of the $gg^{PP}\to b\bar b$
Born cross section is a consequence of Eqs.\ (\ref{eq:a2}) and
(\ref{eq:a3}), and, as discussed, it plays a critical role in controlling
the $b \bar b$ background.
However, as was pointed out already in \cite{borden}, the
suppression of the $J_z=0$ background cross section is removed by the
presence of an additional gluon in the final state. The radiative three-jet
processes can then mimic the two-jet events in some
specific (for instance, quasi-collinear)
configurations. Additional gluon radiation from the `hard' subprocess, $gg^{PP}\to q\bar q$, was considered in detail in \cite{krs1}.

Until now, one source of radiative QCD background has not been addressed
at the quantitative level. This concerns the
contribution from the  $b \bar b g$ events
caused by the `internal' gluon radiation from the spectator
screening gluon, see Fig.~\ref{fig:SG}(b).
Note that a separation between this `internal bremsstrahlung' 
and the `standard' gluon radiation from the active gluons and final
quarks
makes sense only when the momentum $p_t$ of the emitted gluon momenta
exceeds the internal transverse momentum 
in the gluon loop $Q_t$, since at lower momenta (larger wavelengths)
there are interference effects between the different contributions
leading to complete cancellation between radiation from
the $t$-channel gluons for $p_t \ll Q_t$.

Potentially this internal bremsstrahlung contribution
could be quite sizeable. However, as discussed
below, it appears to be numerically
small, in particular, because of an additional suppression resulting from symmetry
arguments, see \cite{myth}. Below, the
detailed evaluation of this background source is presented for the
first time.

\section{Gluon radiation from the screening gluon
 \label{sec:ampl}}
\begin{figure}
\begin{center}
\includegraphics[height=6cm]{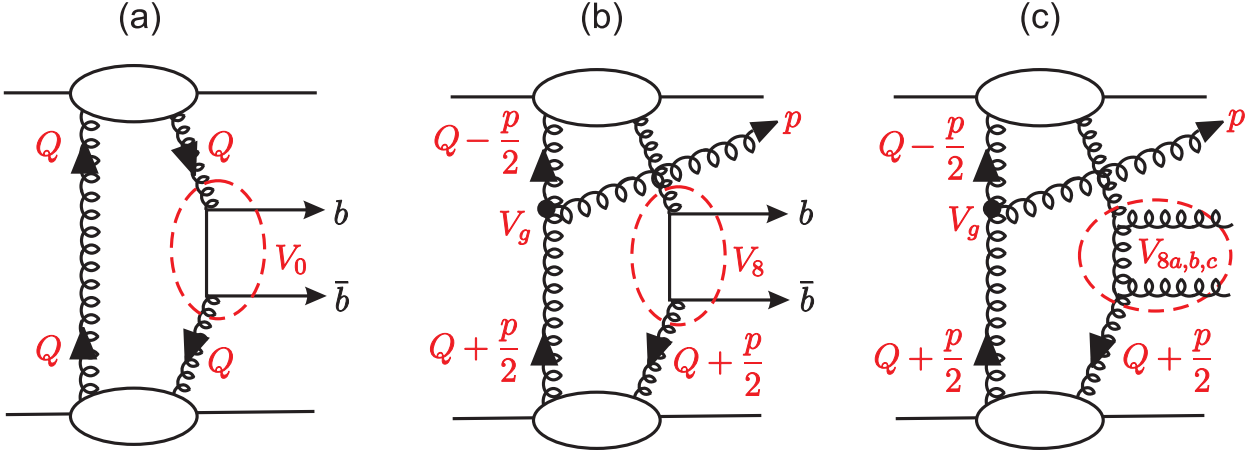}
\caption{Schematic diagrams for (a) $b\bar{b}$, (b) $b\bar{b}+g$ and (c) $gg+g$ exclusive production, where in (b) and (c) the extra gluon, of 4-momentum $p$, is radiated from the screening gluon.}
\label{fig:SG}
\end{center}
\end{figure}

First, recall \cite{KMRprosp} the form of the cross section for non-radiative dijet ($b\bar{b}$) production, that is, for $pp \to p+b\bar{b}+p$ of Fig.~\ref{fig:SG}(a). Similar to the expression for the Higgs signal, (\ref{eq:d1}), the cross section for the QCD background is given by
\begin{equation}
\sigma (pp \to p+b\bar{b}+p) ~\sim~\frac{\langle S^2 \rangle}{B^2} \left| \int \frac{dQ_t^2~V_0}{Q^6_t}f_g(x_1,x'_1,Q^2_t,\mu^2)f_g(x_2,x'_2,Q^2_t,\mu^2) \right|^2~ \hat{\sigma}(gg^{PP} \to b\bar{b}),
\label{eq:V0}
\end{equation}
where the expression in front of $\hat\sigma$ plays the role of the corresponding gluon-gluon luminosity. In other words,
the CEP cross section can be written as
 \begin{equation}
\frac{d\sigma^{\rm CEP}}{dy~d\ln M^2}=
\frac{dL}{dy~d\ln M^2}\cdot \hat\sigma,
\label{eq:CEP}
\end{equation}
where $M$ and $y$ are the mass and rapidity of the centrally
produced system. 

The factor $V_0$ in (\ref{eq:V0}) is the part
of the `hard' matrix element arising from the 
 polarization vectors of the active gluons, which depend on their momenta
$Q$.
For CEP of dijets, the final system
is in a colour-singlet state, and the corresponding 
 factor is just $V_0=Q_t^2$.
 
However, we are interested in radiation from the screening gluon, see Fig.~\ref{fig:SG}(b). Now  we have two hard matrix 
elements - the emission of a new gluon in a left part of the diagram and high $E_T$ dijet production in the right part.
 Therefore the cross section  takes a more complicated form
$$ \sigma=L^{\rm eff}\cdot\hat\sigma$$
with
\begin{equation}
\frac{dL^{\rm eff}}{dy~d\ln M^2}~=~\frac{dL}
{dy~d\ln M^2}~\frac{N_c\alpha_s(p^2_t)/\pi}
{d\ln p^2_t~d\eta_3},
\label{Leff}
\end{equation}
where we have already included in (\ref{Leff}) the soft bremsstrahlung factor $N_c\alpha_s dp^2_t/\pi p^2_t$. That is, using the BFKL effective gluon vertex, it is still possible to write the cross section in a factorized form with an {\it effective} luminosity $L^{\rm eff}$ which now depends on the gluon transverse momentum $p_t$; 
$\eta_3$ is the rapidity of the bremsstrahlung gluon with  momentum $p$.

There are now four propagators (external
to the `hard' matrix element) associated with four gluons with momenta $Q\pm p/2$. 
Calculating the first factor $dL/dyd\ln M^2$ in (\ref{Leff})
we, therefore, have to replace the term
$\int dQ_t^2 V_0/Q_t^6$ in (\ref{eq:V0}) by
\begin{equation}
\int\frac{d^2Q_t}\pi\frac{V_g V_8}{(Q-p/2)^4_t (Q+p/2)^4_t}
\label{denom}
\end{equation}
and, correspondingly, evaluate the unintegrated gluon densities
$f_g$ at $q_t=(Q-p/2)_t$ and $q_t=(Q+p/2)_t$.
Here $p$ is the momentum of the emitted gluon, while $Q_t$ is the momentum of integration
around the gluon loop.
 For exclusive three-jet ($b\bar{b}g$) production, 
the high $E_T$ dijet ($b\bar{b}$) system is in a colour-octet state.
Thus we have denoted the corresponding polarization factor by $V_8$. 
We shall calculate it explicitly later.

Finally, $V_g$, in (\ref{denom}), is the vertex factor arising from the emission of the
extra gluon. It accounts not only for the diagrams shown in 
Fig.~\ref{fig:SG}(b,c), but also for soft gluon radiation from the upper and lower blobs as well. The simplest way to calculate $V_g$ is to
use the effective
vertex of gluon emission in the BFKL approach \cite{BFKLvg}, see also \cite{BFKL}.
We choose the frame where the emitted gluon rapidity $\eta_3=0$, and direct the axis $x$ 
along the gluon transverse momentum  $\vec p_t$.
In this case the gluon polarization has two transverse components,
$z$ and $y$, with vertex factors
 \begin{equation}
C_z=2(Q-p/2)_t\cdot(Q+p/2)_t/|p|,
\label{vertex-z}
\end{equation}
\begin{equation}
C_y=2Q_y,
\label{vertex-y}
\end{equation}
for the gluons polarized in the $z$ and $y$ directions
respectively.

 For the CEP of the  $b\bar b g$-system  we can neglect the term which is linear in $Q_t$. 
This term vanishes after angular integration,
 since the $Q_t$-component directed along the gluon momentum $\vec{p}_t$ 
is orthogonal to the gluon polarization vector $e^\mu_3$, while the whole expression
 is symmetric under the 
interchange $Q_y\to-Q_y$.
 Thus, with the help of 
the effective BFKL vertex 
we obtain
\begin{equation}
V_g=(Q-p/2)_t(Q+p/2)_t=(Q_t^2-p^2_t/4).
\label{vg}
\end{equation}

To calculate
the $V_8$ factor, 
we can either use the Weizs\"{a}cker-Williams method or choose
a suitable (planar) gauge. Then, to LO accuracy (both in the collinear, $\ln Q^2$, and  BFKL, $\ln(1/x)$, approaches), we can replace
the polarization vectors $e^\mu_i$ of the $t$-channel active gluons (of momenta $q_i=Q\pm p/2$)  by their
transverse momenta. Explicitly, we have $e^\mu_i\simeq q^\mu_{it}/x_i$. The factor
$1/x_i$ is included in the unintegrated gluon densities $f_g$.  
As a result, in the calculation of 
the hard matrix element, we shall write the polarization tensor
corresponding to the incoming active gluons, in the form
\begin{equation}
e^\mu_1e^\nu_2~\propto~T^{\mu\nu}= 
\left[\left(Q-\frac p2\right)^\mu\left(Q+\frac p2\right)^\nu\right]_t.
\label{tmunu}
\end{equation}

We are now ready to compute the cross sections for, first, the subprocess $gg \to b\bar{b}+g$, and then for $gg \to gg+g$, in which a third jet, $g$, is emitted.

Note that, because of the gluon vertex factors (\ref{vertex-z},\ref{vertex-y}),
 the integral (\ref{denom}) for the effective luminosity $L^{\rm eff}$ for gluon bremsstrahlung from the screening gluon is less sensitive to the infrared region than the analogous integral (\ref{eq:V0})
for the standard exclusive luminosity $L$ in the non-radiative case.

\section{Exclusive $b\bar b +g$ production}
\label{sec:bbg}

\begin{figure}
\begin{center}
\includegraphics[height=5cm]{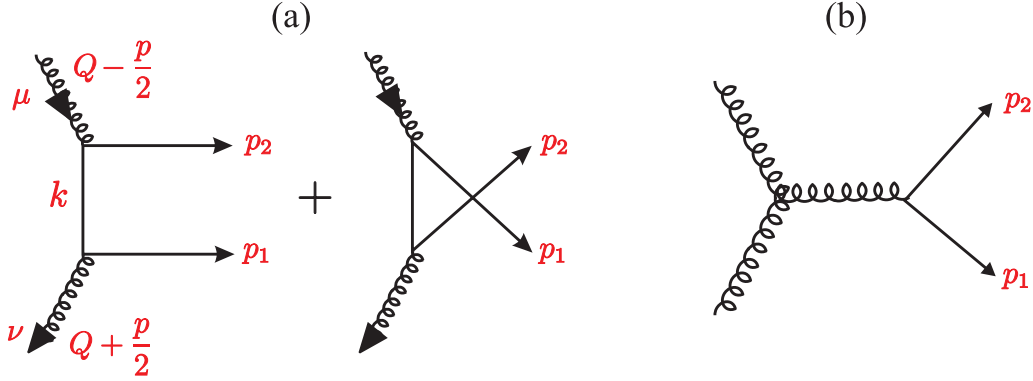}
\caption{Diagrams for the $gg\to b\bar b$ subprocess.}
\label{fig:VF2}
\end{center}
\end{figure}

First, recall that the $x$ axis is directed along the emitted gluon transverse momentum $\vec p_t$ and note, that the component of the tensor 
$T^{\mu\nu}$ linear in $Q_y$, that is $Q_yp_x-p_xQ_y$,
vanishes for the $gg\to b\bar b$ matrix element in the massless quark limit.
Indeed, such a component corresponds to the $J_z=0$,  
octet state of the incoming gluons,
for which the $gg\to b\bar b$ matrix element vanishes in the massless quark limit,
see footnote 5.
Therefore, we need consider only the polarization tensor
\be
e^\mu_1e^\nu_2~\propto~T^{\mu\nu}=(Q-p/2)_x(Q+p/2)_x+Q_yQ_y,
\ee
where the indices $xx$ and $yy$ on the right-hand-side play the role of
 $\mu\nu$ on the left-hand-side.
 The contributions to the 
$gg\to b\bar b$ amplitude, corresponding to Fig.~\ref{fig:VF2}(a), contain $\bar u(p_1)e\!\!\! /^\nu k\!\!\! /e\!\!\! /^\mu v(p_2)$,
 where $k$ is the $t$-channel quark momentum.
The sum of the contributions with a longitudinal component of $k$ and the contribution of the diagram with $s$-channel gluon, Fig.~\ref{fig:VF2}(b), vanishes in the massless quark limit, analogously to
the $J_z=0$ case.
The $T^{xx}$ component gives $ \gamma_xk\!\!\! /_t\gamma_x=q\!\!\! /$ where $\vec q=\vec k_x-\vec k_y$,
that is $q_x=k_x$ and $q_y=-k_y$. 
Similarly, the $T^{yy}$ component
gives $-q\!\!\! /$. Thus, when calculating  the effective $gg$ luminosity for {\it octet} $q\bar q$ production 
 we have to use 
\begin{equation}
V_8~\equiv~V_{8q}~=~
\left[\left(Q-\frac p2\right)_x\left(Q+\frac p2\right)_x-Q_yQ_y\right],
\label{v8q}
\end{equation}
and then to calculate the square of the matrix element, that is Tr$[q\!\!\! /p\!\!\! /_1
q\!\!\! /p\!\!\! /_2]$.
The resulting octet $b\bar b$ production is  
\begin{equation}
\frac{d\hat\sigma(b\bar b)_8}{dt}=\frac 3{64}
\frac{\pi\alpha_s^2(E^2_T)}{M^4}
\left(\frac{{\rm cos}^2\theta((1+{\rm cos}^2\theta)+{\rm sin}^2\theta(1-2{\rm cos}^2(2\phi)))}
{{\rm sin}^2\theta}\right).
\label{sig8q}
\end{equation}
The coefficient  $3/64$ accounts for the colour factors, arising both
in the $gg\to b\bar b$ subprocess and in the radiation of the additional gluon. In (\ref{sig8q}), $\theta$ is the polar angle of a high $E_T$
 jet in the frame where the rapidity of high-$E_T$ dijet system $y_{b \bar b}=0$, and 
$\phi$ is the azimuthal angle between the gluon momentum $p$ (i.e. $x$ axis) and the quark jet direction. After integration over this azimuthal angle the second
term in the brackets in the numerator of (\ref{sig8q})
disappears, since $\langle 1-2{\rm cos}^2(2\phi)\rangle=0$ in the limit  $E_T\gg p_t$.

Instead of  working in terms of the dijet variables $E_T$ and
$\delta\eta=\eta_1-\eta_2$ ($\delta\eta$ is the rapidity difference between
the two outgoing jets)  here, following \cite{KMRprosp} we use the subprocess variables $M^2$ and $t$. It turns out that factors in Jacobian cancel so that we obtain
$$\frac{d\hat\sigma}{dt}\left/\frac{dM^2}{M^2}\;=\;\frac{d\hat\sigma}{dE^2_T}\right/{d(\delta\eta)}\; .$$
Thus, the product of the luminosity $dL/dyd\ln M^2$ and $d\hat\sigma/dt$ gives the differential
cross section $d\sigma/dyd(\delta\eta)dE^2_T$.

Note that the integrated cross section vanishes at 90$^o$ when ${\rm cos}^2\theta=0$.
This is a general property of the QHC  $gg\to b\bar b$ amplitudes
with  $J_z=2$, see \cite{krs1,myth}.
In order to gain insight into the origin of
such a suppression, we note that when $\theta=\pi/2$ in the $b \bar b$ rest frame, the final
state is asymmetric with respect
 to a $180^\circ$ rotation
about the quark direction: the overall spin projection onto this direction
is $\pm 1$. At the same time, the initial digluon system is symmetric
under this rotation due to the identity
of the incoming gluons (protons). This symmetry, in configuration space,
is not affected by the colour structure, and, as a result of rotational invariance,
the $\theta=\pi/2$ amplitude vanishes \cite{krs1,borden}.
As discussed in Ref.\cite {krs1}, since we are interested in
 the detection of large angle jets
in the central detectors at the LHC, the fact that all  $J_z=2$ QHC
cross sections are proportional to ${\rm cos}^2\theta$ provides an
additional suppression of all such background contributions
(by a factor of, at least,  $\sim 0.2$).
This suppression provides an improvement of the ${b \bar b}$ signal-to-background.

It is worth mentioning, that,
strictly speaking, we have to account
for the interference between the radiation from the screening gluon
and that  from the active gluons and from the final quarks (on the
right-hand side of the amplitude in Fig.~\ref{fig:SG}(b)).
However, in reality, the amplitude
squared, $|A_{\rm scr}|^2$, corresponding to the emission from
the screening 
gluon is so small that the interference term (which is suppressed as the 
first power of $A_{\rm scr}$) is still negligible.
Therefore, here we evaluate
only the contribution 
 of radiation from the screening gluon and do
not consider a more complicated  expression for the 
interference. To illustrate that the amplitude $A_{\rm scr}$ 
is small we note that for
 $p_t\ll Q_t$ the polarization tensor  $T^{\mu\nu}$ is proportional $(Q^\mu Q^\nu)_t$.
Thus after the azimuthal integration, the initial digluon system is in a  $J_z=0$ state. Therefore,
 the corresponding matrix element for the octet $gg\to b\bar b$ hard subprocess is suppressed 
by the $J_z=0$ selection rule. The same is true for 
 emission off the external quarks
in the limit $p_t\ll E_T$.
This is also 
 suppressed due to the $J_z=0$ selection rule.
At the same time, for relatively large gluon momenta,
 $p_t\gapproxeq 2Q_t$, the effective luminosity for exclusive $b\bar b g$ production
is suppressed by a large factor, $(Q/p)^4$, as compared
to the CEP of exclusive dijets.
Therefore, we do not need to worry about the explicit calculations
of these type of interference.

Moreover, at LHC energies, the typical value of the loop integration momentum 
is $Q_t\sim 2$ GeV. Now a gluon with $p_t>2Q_t$, that is a gluon with $p_t\gapproxeq 5$ GeV, in a low multiplicity exclusive process should be {\it observed} experimentally as an individual minijet. Thus such a process
should not be considered as a background to exclusive 
$H\to b\bar b$ production.

\section{Exclusive $gg+g$ production}
\label{sec:bbg}
Here
 we present  analytical results for gluon radiation from the screening gluon, which accompanies a hard  $gg\to gg$  
process. Recall that exclusive $gg \to ggg$ production can be a background to exclusive $b\bar b$ production, since gluon jets can mimic $b$ jets. To describe the three-gluon exclusive production process of Fig.~\ref{fig:SG}(c)
 we  take the same $V_g$
vertices, (\ref{vertex-z},\ref{vertex-y}), describing the extra gluon emission,
 and the same  polarization tensor, $T^{\mu\nu}$ of the active gluons,
 (\ref{tmunu}), and  calculate the corresponding
 $gg\to gg$ amplitude of the hard subprocess,
where now the two incoming `active' gluons are in a {\it colour-octet} state. 

As in Ref.~\cite{krs1}, it is convenient to use the helicity formalism of \cite{MP}.
\begin{figure}
\begin{center}
\includegraphics[height=5cm]{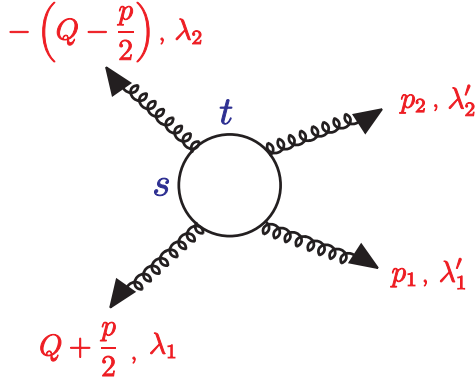}
\caption{Momenta and helicities of the $gg\to gg$ subprocess.}
\label{fig:VF3}
\end{center}
\end{figure}
The non-zero amplitudes for three basic helicity configurations are
$$(++;--)=(--;++)=C's^2/tu$$
$$(+-;+-)=(-+;-+)=C't^2/tu$$
\begin{equation}
(+-;-+)=(-+;+-)=C'u^2/tu,
\label{ggmhv}
\end{equation}
where the  notation $(\lambda_1\lambda_2;\lambda'_1\lambda'_2)$= $(++;--)$, etc.
 corresponds to the helicities
of the gluons as defined in  Fig.~\ref{fig:VF3}. The normalization 
factor is $C'=8\pi\alpha_s$, up to a colour coefficient. Here
 $s,t$ and $u$ are the standard Mandelstam variables for
 hard $gg\to gg$ scattering.

In terms of helicity, the elements of the polarization tensor of the active gluons,  
(\ref{tmunu}), 
can be written as
$$ (xx+yy)/2=(++)+(--)\;\;\;\;\;\;\;\;\;\;\;(a)$$
$$ (xx-yy)/2=(+-)+(-+)\;\;\;\;\;\;\;\;\;\;\;(b)$$
$$ i(xy-yx)/2=(++)-(--)\;\;\;\;\;\;\;\;\;\;\;(c)$$
\begin{equation}
i(xy+yx)/2=(+-)-(-+)\;\;\;\;\;\;\;\;\;\;\;(d)
\label{xx-xy}
\end{equation}
Thus, we have to compute the coherent sum of the amplitudes $A_{ji}$ corresponding to the 
$T^{\mu\nu}$ tensor in initial state for each helicity configuration $(ji)$ in the final state,
and then
to take the sum $\sum_{ji}|A_{ji}|^2$. Recall that
the initial states $(++)$,$(--)$,$(-+)$,$(+-)$ correspond  to the di-gluon $J_z$ projection
equal to 0, 0, +2 and $-2$ respectively. Obviously, 
these states do not interfere with each other. 

Again we neglect the interference with
 amplitudes where the third gluon is radiated from the hard subprocess.
Therefore, we have to calculate the effective luminosities\footnote{The effective luminosities are defined in eq. (\ref{Leff}).}, which contain the soft bremsstrahlung factor for the extra gluon. $L^a$, $L^b,\ L^c$
for the initial polarization states  $(a,b,c)$. Note that for $(d)$
the corresponding luminosity, $L^d$, is very small: in terms of the
 $T^{\mu\nu}$
 tensor, (\ref{tmunu}),  $L^d$ results from the $Q_xQ_y+Q_yQ_x$ component,
which vanishes due to $Q_x\leftrightarrow -Q_x$ symmetry\footnote{This symmetry
could be violated by a different scale $q_t^2$
 behaviour of the unintegrated gluons $f_g(x_i,x'_i,q_t^2,\mu^2)$
at different values of $x_1$ and $x_2$, 
but this effect is negligibly small.}.
 
The effective luminosities  $L^a(0),\ L^b(2)$, which correspond to $J_z=0, 2$,
can be calculated using the vertex factor $V_g$ of (\ref{vg}), and using 
the combinations which give the appropriate $V_8$ factors
\begin{equation}
V_{8a}=(Q^2-p^2/8)\;\;\;\;\mbox{and }\;\;\;\; V_{8b}=-p^2/8.
\label{lab}
\end{equation}
Similarly, the luminosity $L^c(0)$, which corresponds to $J_z=0$, can be obtained
using the vertex factor 
 $V_g=Q_y$  and
\begin{equation}
V_{8c}~=~|Q|~|p|.
\label{lc}
\end{equation}
Finally, these luminosities should be multiplied by the
 corresponding $J_z=0,2$ cross sections for $s$-channel {\it octet} $gg\to gg$ hard scattering.
We find
\begin{equation}
\frac{d\sigma^{(0)}}{dt}~=~C~\frac{\pi\alpha^2_s}{E^4_T}
\label{sig0}
\end{equation}
and
\begin{equation}
\frac{d\sigma^{(2)}}{dt}~=~C~\frac{\pi\alpha^2_s}{E^4_T}
~{\rm cos}^2\theta~,
\label{sig2}
\end{equation}
where the coefficient $C= 9/16$ incorporates all colour factors.

\section{Numerical results and discussion}
\label{sec:disc}

\begin{figure} [t]
\begin{center}
\includegraphics[height=10cm]{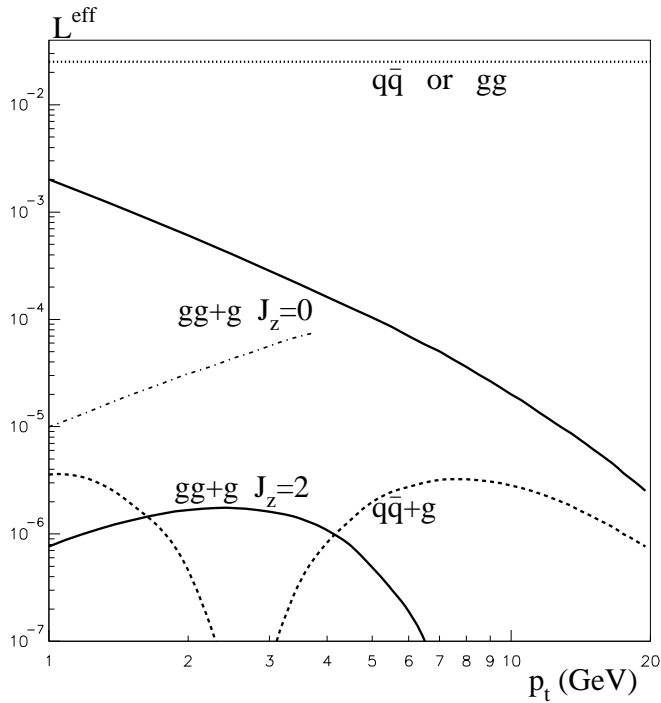}
\caption{The effective luminosities, $L^{\rm eff}$ of (\ref{Leff}), versus
transverse momentum, $p_t$ of the bremsstrahlung gluon, for the colour-octet radiative processes ($pp \to p+q\bar{q}g+p$ and $pp\to p+ggg+p$) for $\sqrt{s}=14$ TeV, $y=0$  
and $M=120$ GeV.
The curves are  for the case of $\eta_3=0$, however, as far as the bremsstrahlung 
gluon energy does not drastically affect the overall kinematics
 the results are independent on $\eta_3$. 
 In this plot we use the notation $q\bar{q}$, as it is sufficient to treat the $b$ quark here as massless.}
\label{fig:Leff}
\end{center}
\end{figure}
For illustration, we consider the exclusive production of a central system of mass $M=120$ GeV at rapidity $y=0$ and LHC energy $\sqrt{s}=14$ TeV. The effective luminosities\footnote{Note,
 that the soft gap survival factor $S^2$ is not included in the luminosities, $L^{\rm eff}$, which are shown in Fig.~\ref{fig:Leff}.}, $L^{\rm eff}$ of (\ref{Leff}), of exclusive processes with radiation from the screening gluon are shown in Fig.~\ref{fig:Leff}.

\subsection{Results for exclusive $b\bar{b}+g$ production}


First, we consider the effect of  gluon radiation (from the screening gluon) which  accompanies exclusive  $b\bar b$ production, as shown in  Fig.~\ref{fig:SG}(b).
 Recall that exclusive QCD $b\bar b$ production is the main irreducible background to CEP of a Higgs boson which decays via
 the $H\to b\bar b$  mode. In the absence of
 gluon radiation, the $J_z=0$ selection rule \cite{KMRmm},
suppresses the QCD production of the quark-antiquark pair by a factor of $m^2_b/M^2$. 
Since the radiation of an additional gluon does not obey the selection rule,
the $b\bar b +g$ background (which is not suppressed by the 
$m^2_b/M^2$ ratio\footnote{Since we are looking for the contribution 
which is not suppressed by the small $m^2_b/M^2$ factor, it is sufficient
 to calculate the corresponding cross section treating the $b$ quark as massless.})
could potentially  become
comparable with the original LO background.
 However, this is not the case. 
In Fig.~\ref{fig:Leff} the dotted line, marked $q\bar q$ (or $gg$), shows the luminosity
 which is to be convoluted with the LO binary $gg\to b\bar b$ (or $gg\to gg$) colour singlet 
cross section. On the other hand, the effect of radiation of an extra gluon from the screening $t$-channel gluon is evaluated as the product of $L^{\rm eff}$ (shown by the dashed line marked $b\bar b +g$)\footnote{Note, that
the origin of the dip in the region $p_t\sim 2.5$ GeV is just a consequence of gauge invariance, which leads to
the factor $(Q-p/2)_t(Q+p/2)_t$ in equations (\ref{vg},\ref{tmunu}).}
 by the octet cross section (\ref{sig8q}).
We see that this luminosity 
(dashed line) is about four orders  of magnitude lower than that for binary $b\bar b$
 exclusive production (dotted line)\footnote{As  mentioned in Section 4, the expression 
 for the effective luminosity $L^{\rm eff}$ is less sensitive to the infrared region than the analogous 
result for the standard exclusive luminosity $L$ corresponding to the non-radiative case. 
To quantify the role of the contribution coming from  low values of $(Q_t\pm p_t/2)$, we calculate
 the value of $L^{\rm eff}$ using different infrared cutoffs. At $p_t=6$ GeV the cutoff 
of $Q_t\pm p_t/2 > 1$ GeV decreases the amplitude of exclusive $b\bar b+g$ production 
(with gluon emission from the screening gluon) by about 20 \% . On the other hand, since the factor $V_{8q}$ in (\ref{v8q}) can change sign, in the dip region ($p_t=2\ -\ 3$ GeV) the amplitude becomes even larger than without the cutoff (but is still small).
In any case, the contribution coming from the infrared region 
is beyond the perturbative QCD framework. Therefore,
when calculating the curves in Fig. 5 we introduce our standard infrared cutoff
of 0.85 GeV. }. 

Note that the cross section of the hard-octet subprocess (\ref{sig8q}) 
 vanishes at $90^o$. After the integration over an appropriate angular interval, $60^o<\theta <120^o$, we find that the ratio of the octet $b\bar b$  and the LO singlet $b\bar b$ subprocess cross sections is 
\be
\frac{\hat\sigma(b\bar b)_8}{\hat\sigma(b\bar b)_0}~=~4.7.
\label{rat}
\ee
 Recall that the LO cross section, $\hat\sigma(b\bar b)_0$, 
is suppressed by a factor of $m^2_b/E^2_T$. The NLO
correction to $\hat\sigma(b\bar b)_0$ \cite{shuv} reduces the cross section by a further factor of about 2, and so increases the ratio (\ref{rat}) by this factor.
 However, when we account for the huge difference of 
the corresponding luminosities we see that,
 despite the large ratio 
$\hat\sigma(b\bar b)_8/\hat\sigma(b\bar b)_0$, the background caused by radiation from the screening gluon is negligible in comparison with the irreducible background due to exclusive $b\bar b$ production.

Up to now we discuss just the case when the relatively soft gluon in exclusive $b\bar b+g$ production is emitted from the screening (left in Fig.2) gluon. Since the corresponding amplitude is very small a larger contribution may come from the interference between the gluon emission from the hard matrix element and the screening gluon. On the other hand, as we have seen, the probability of the octet $b\bar b$ pair production accompanished by the emission from the screening gluon is more than a factor $|F|^2\sim 10^{-4}$ (four orders of magnitude) smaller than the probability of exclusive singlet $b\bar b$ production. The coresponding interference contribution should be suppressed at least by $F\sim 10^{-2}$ and thus will be also negligible in comparison with the irreducible $b\bar b$ background.

\subsection{Results for exclusive $gg+g$ production}

Now we turn to exclusive digluon production. The two continuous curves in Fig.~\ref{fig:Leff} show the effective luminosities for gluon dijet production accompanied by an extra gluon radiated from the screening gluon, see Fig.~\ref{fig:SG}(c). The difference in these luminosities reflects the 
difference in the
 quantum numbers of the active incoming di-gluon system. The upper curve corresponds to $J_z=0$, while the lower ($J_z=2$) curve
 is suppressed by $s$-channel helicity conservation
\cite{KMRmm}; in the limit of 
$p_t\to 0$ we return to the $J_z=0$ selection rule.
It is worth mentioning that the difference  between the `$gg+g$' and the `$q\bar q+g$'
effective luminosities, which both correspond to $J_z=2$, follows from the
fact that the large-angle $gg\to b\bar b$ QHC amplitude in the massless limit
should be $P$-even, and thus, the contribution generated by the term   (\ref{vertex-y})
vanishes.

When evaluating the physical $b\bar b$ background contribution caused
by the $gg+g$ events, the corresponding $gg\to gg$
cross section should be multiplied by a very small factor, $P^2_{g/b}$,
see \cite{DKMOR}; where $P_{g/b}\sim 0.01$ 
is the probability  of misidentification of a gluon jet as a $b$-quark jet.
As a result, this background contribution also becomes numerically
negligibly small as
compared to the main irreducible background, 
$gg^{PP}\to b\bar b$.

Finally, we note that, at first sight,
the $J_z=0$ contribution at very low $p_t$ is quite large.
The corresponding effective luminosity, shown by a solid line, is only an order 
of magnitude lower than that for two-gluon CEP production.
However, as already discussed, at such low
$p_t$, it is necessary 
to include the destructive interference with soft-gluon emission in the matrix element of the hard subprocess.
Accounting for these  interference effects, we arrive at the final 
result shown by the dot-dashed curve in  Fig.~\ref{fig:Leff}. 
To calculate the amplitude of soft-gluon  emission  (within the  $d\ln p^2_t~d\eta_3$
element of phase space in the beam direction) from the `hard' matrix element, we  have 
to multiply the known amplitude of exclusive gluon dijet production~\cite{KMRprosp} by the soft bremsstrahlung factor  $\sqrt{N_c\alpha_s(p^2_t)/\pi}$
 \footnote{We do not continue the dot-dashed curve into the 
domain of larger $p_t$ since: i) a gluon with larger $p_t$ can be observed experimentally as a separate jet,
  ii) for  larger $p_t$ a more precise and complicated
expression for the radiation from the hard matrix element should be used.}.

\section{Summary}

One of the advantages of exclusive processes is that the 
major irreducible QCD $b\bar b$ background is strongly suppressed by 
a $J_z=0$ selection rule, which leads to a factor $m^2_b/E^2_T$ in the CEP cross section. This offers the unique 
possibility to observe an exclusive Higgs boson in its main decay 
mode, $H\to b\bar b$.
However, the suppression can be removed by the presence of an additional gluon in the final state. The radiation from the matrix element of a `hard' subprocess was considered in \cite{krs1}.
Here, we have explicitly calculated the effect of radiation from the screening gluon for the case of $b\bar b+g$ and $gg+g$ final states,
 where the $b\bar b$ or $gg$ dijets are in a colour-octet state. (The $gg+g$ configuration was studied to allow for the possibility that gluon jets may be misidentified as $b$-quark jets.)
 We found that these channels give a completely negligible
contribution to QCD background to exclusive $pp\to p+(H\to b\bar b)+p$ production at the LHC; much less than the already suppressed
exclusive $b\bar b$ production, $pp\to p+(b\bar b)+p$.

\section*{Acknowledgements}
We thank Albert De Roeck, Risto Orava, Andy Pilkington, Marek Tasevsky
and James Stirling 
for useful discussions. 
MGR thanks the IPPP at the University of Durham for hospitality.
The work was supported by  grant RFBR
07-02-00023, by the Russian State grant RSGSS-3628.2008.2.


\begin{thebibliography}{99}

\bibitem{ILC}
G.~Weiglein {\it et al.},  (LHC/LC Study Group Collaboration),
{\tt arXiv:hep-ph/0410364};\\
S.~Heinemeyer,
  Acta Phys.\ Polon.\  B {\bf 39}, 2673 (2008)
  [arXiv:0807.2514 [hep-ph]];\\
A.~Djouadi and R.M.~Godbole,
  arXiv:0901.2030 [hep-ph],
and references therein.
\bibitem{FP420}
M.G.~Albrow {\it et al.},  [FP420 R\&D Collaboration],
arXiv:0806.0302 [hep-ex].
\bibitem{pb} P.J.~Bussey,
 arXiv:0809.1335 [hep-ex];\\
  M.~Grothe  [CMS Collaboration],
arXiv:0806.2977 [hep-ex];\\
D.~d'Enterria,
  arXiv:0905.4307 [hep-ex].
%
%
\bibitem{KMRprosp} V.A. Khoze, A.D. Martin, M.G. Ryskin, Eur. Phys. J. C {\bf 23}, 311 (2002).
\bibitem{DKMOR} A.~De~Roeck {\it et al.}
Eur. Phys. J. C {\bf 25}, 391 (2002).
\bibitem{hkrstw}S.~Heinemeyer {\it et al.}, 
  Eur.\ Phys.\ J.\  C {\bf 53}, 231 (2008)
  [arXiv:0708.3052 [hep-ph]].
\bibitem{hkrtw} S.~Heinemeyer {\it et al.}
  arXiv:0811.4571 [hep-ph].
\bibitem{cox1}
  B.E.~Cox, F.K.~Loebinger and A.D.~Pilkington,
  JHEP {\bf 0710}, 090 (2007) 
  [arXiv:0709.3035 [hep-ph]].

                                                             
\bibitem{KKMRext} A.~Kaidalov {\it et al.},
                  Eur. Phys. J. C {\bf  33}, 261 (2004),
                  hep-ph/0311023.
 \bibitem{JE}
  J.R.~Ellis, J.S.~Lee and A.~Pilaftsis,
  Phys.\ Rev.\  D {\bf 71}, 075007 (2005)
  [arXiv:hep-ph/0502251].
                  
 \bibitem{fghpp} J.R.~Forshaw {\it et al.},  
  JHEP {\bf 0804}, 090 (2008)
  [arXiv:0712.3510 [hep-ph]].
\bibitem{royon} C.~Royon,
  Acta Phys.\ Polon.\  B {\bf 39}, 2339 (2008)
  [arXiv:0805.0261 [hep-ph]].
 
\bibitem{katri}M.~Chaichian {\it et al.},
  arXiv:0901.3746 [hep-ph], JHEP in press.

 \bibitem{krs1}   V.A.~Khoze, M.G.~Ryskin and W.J.~Stirling,
  Eur.\ Phys.\ J.\  C {\bf 48}, 477 (2006) 
  [arXiv:hep-ph/0607134].

\bibitem{KMRins} V.A.~Khoze, A.D.~Martin and M.G.~Ryskin,
  Phys.\ Lett.\  B {\bf 650}, 41 (2007)
  [arXiv:hep-ph/0702213].

\bibitem{shuv} A.G.~Shuvaev, 
 V.A.~Khoze, A.D.~Martin and M.G.~Ryskin,
  Eur.\ Phys.\ J.\  C {\bf 56}, 467 (2008)
  [arXiv:0806.1447 [hep-ph]].

\bibitem{Blois07} V.A.~Khoze, M.G.~Ryskin and A.D.~Martin, in
{\it `Hamburg 2007, Blois07, Forward physics and QCD'}, p.452-458.


\bibitem{KMRmm}  V.A.~Khoze, A.D.~Martin and M.G.~Ryskin,
  Eur.\ Phys.\ J.\  C {\bf 19}, 477 (2001)
  [Erratum-ibid.\  C {\bf 20}, 599 (2001)]
  [arXiv:hep-ph/0011393].

\bibitem{KKMRcentr} A.B.~Kaidalov {\it et al.},
  Eur.\ Phys.\ J.\  C {\bf 31}, 387 (2003)
  [arXiv:hep-ph/0307064].
%
\bibitem{KMRCP} V.A.~Khoze, A.D.~Martin and M.G.~Ryskin, Eur. Phys. J. C {\bf 34}, 327 (2004).
%
%
\bibitem{KMR} V.A. Khoze, A.D. Martin, M.G. Ryskin,
Eur.\ Phys.\ J.\  C {\bf 14}, 525 (2000) 
  [arXiv:hep-ph/0002072].

\bibitem{KMRsoft} 
V.A.~Khoze, A.D.~Martin and M.G.~Ryskin,
  Eur.\ Phys.\ J.\  C {\bf 18}, 167 (2000) 
  [arXiv:hep-ph/0007359].
\bibitem{KMRnewsoft} M.G. Ryskin, A.D. Martin and V.A.~Khoze, Eur. Phys. J. C {\bf 54}, 199
(2008);\\
E.G.S. Luna {\it et al.}, Eur. Phys. J. C {\bf 59}, 1
(2009), arXiv:0807.4115 [hep-ph].

\bibitem{KMRnns} M.G.~Ryskin, A.D.~Martin and V.A.~Khoze,
  Eur.\ Phys.\ J.\  C {\bf 60}, 265 (2009)
  [arXiv:0812.2413 [hep-ph]].
\bibitem{KMRearly} V.A.~Khoze, A.D.~Martin and M.G.~Ryskin,
Eur.\ Phys.\ J.\  C {\bf 55}, 363 (2008) 
  [arXiv:0802.0177 [hep-ph]].

\bibitem{MRKepip} A.D.~Martin, M.G.~Ryskin and V.A.~Khoze, Acta Phys. Pol. B {\bf 40}, 1841 (2009)
  [arXiv:0903.2980 [hep-ph]].
\bibitem{KMRS}
V.A.~Khoze, A.D.~Martin, M.G.~Ryskin and W.J.~Stirling,
Eur.\ Phys.\ J.\  C {\bf 35}, 211 (2004) 
  [arXiv:hep-ph/0403218].
\bibitem{CDFdijet} T.~Aaltonen {\it et al.}  [CDF Collaboration],
  Phys.\ Rev.\  {\bf D77}, 052004 (2008).
\bibitem{CDFgg} T.~Aaltonen {\it et al.}  [CDF Collaboration],
  Phys.\ Rev.\ Lett.\  {\bf 99}, 242002 (2007).
\bibitem{CDFchi} T.~Aaltonen {\it et al.}  [CDF Collaboration],  Phys.\ Rev.\ Lett.\ {\bf 102}, 242001 (2009).

\bibitem{JHEP} 
A.B.~Kaidalov, V.A.~Khoze, A.D.~Martin and M.G.~Ryskin,
  Eur.\ Phys.\ J.\  C {\bf 47}, 385 (2006)
  [arXiv:hep-ph/0602215];\\
V.A.~Khoze, A.D.~Martin and M.G.~Ryskin,
  JHEP {\bf 0605}, 036 (2006).

\bibitem{vakjj} V.A.~Khoze, A.D.~Martin and M.G.~Ryskin,
  Frascati Phys.\ Ser.\  {\bf 44}, 147 (2007)
  [arXiv:0705.2314].

\bibitem{Liverpool} V.A. Khoze, A.D. Martin and M.G. Ryskin,
arXiv:hep-ph/0006005, {\it in} Proc. of 8th Int. Workshop on Deep
Inelastic Scattering and QCD (DIS2000), Liverpool, ed. by J.Gracey,
T.Greenshaw (World Scientific, 2001), p.592.

\bibitem{cudell} 
J.R.~Cudell, A. Dechambre, O.F. Hernandez and I.P. Ivanov,
  Eur.\ Phys.\ J.\  C {\bf 61}, 369 (2009) [arXiv:0807.0600].

\bibitem{borden}D.L.~Borden, V.A.~Khoze, W.J.~Stirling and J.~Ohnemus,
  Phys.\ Rev.\   {\bf D50}, 4499 (1994).


\bibitem{fkm} V.S. Fadin, V.A. Khoze and A.D. Martin, Phys. Rev.
{\bf D56} (1997) 484.

\bibitem{mhv1} S.J.~Parke and T.R.~Taylor,
Phys.\ Rev.\ Lett.\  {\bf 56} (1986) 2459;\\
F.A.~Berends and W.T.~Giele,
Nucl.\ Phys.\ {\bf B306} (1988) 759.

\bibitem{MP}
M.L.~Mangano and S.J.~Parke,
Phys.\ Rept.\  {\bf 200}, 301 (1991).

\bibitem{myth}V.A.~Khoze, A.D.~Martin and M.G.~Ryskin,
  Eur.\ Phys.\ J.\  C {\bf 26}, 229 (2002)
  [arXiv:hep-ph/0207313].
\bibitem{BFKLvg}E.N.~Antonov, L.N.~Lipatov, E.A.~Kuraev and I.O.~Cherednikov,
  Nucl.\ Phys.\  B {\bf 721}, 111 (2005)
  [arXiv:hep-ph/0411185].
\bibitem{BFKL}V.S.~Fadin, E.A.~Kuraev, and L.N.~Lipatov,
Phys. Lett. B {\bf60}, 50  (1975); \\
E.A.~Kuraev, L.N.~Lipatov, and V.S.~Fadin, Zh. Eksp. Teor. Fiz.
{\bf 71}, 840 (1976) [Sov. Phys. JETP {\bf 44}, 443 (1976)]; {\it
ibid.} {\bf 72}, 377 (1977) [{\bf 45}, 199 (1977)];\\
I.I.~Balitsky and L.N.~Lipatov, Yad. Fiz. {\bf28}, 1597 (1978)
[Sov. J. Nucl. Phys. {\bf28}, 822 (1978)];\\
L.~N.~Lipatov,
  Phys.\ Rept.\  {\bf 286}, 131 (1997)
  [arXiv:hep-ph/9610276].


%
\end{thebibliography}
\end{document}